\documentclass[conference,10pt,letterpaper,twocolumns,final]{IEEEtran}

\usepackage{cite}

\usepackage{amsmath,amssymb,amsthm,fixmath}

\usepackage{xcolor}

\usepackage[inline]{enumitem}

\usepackage{siunitx}
\DeclareSIUnit{\dBm}{dBm}

\usepackage{graphicx}
\usepackage{subcaption}

\usepackage{epstopdf}
\usepackage{cuted}
\usepackage{multirow,booktabs}
\usepackage{diagbox}

\usepackage{optidef}

\usepackage[hidelinks]{hyperref}

\usepackage[shortcuts,acronym,automake]{glossaries}
\makeglossaries
\newacronym{awgn}{AWGN}{additive white Gaussian noise}
\newacronym{bcd}{BCD}{block coordinate descent}
\newacronym{csi}{CSI}{channel state information}
\newacronym{csit}{CSIT}{channel state information at transmitter}
\newacronym{fbl}{FBL}{finite blocklength}
\newacronym{noma}{NOMA}{non-orthogonal multi-access}
\newacronym{nom}{NOM}{non-orthogonal multiplexing}
\newacronym{ofdm}{OFDM}{orthogonal frequency-division multiplexing}
\newacronym{per}{PER}{packet error rate}
\newacronym{phy}{PHY}{physical}
\newacronym{pls}{PLS}{physical layer security}
\newacronym{prb}{PRB}{physical resource block}
\newacronym{sic}{SIC}{successive interference cancellation}
\newacronym{sinr}{SINR}{signal-to-interference-and-noise ratio}
\newacronym{snr}{SNR}{signal-to-noise ratio}

\usepackage{tcolorbox}
\tcbuselibrary{many}
\newtheorem{theorem}{Theorem}
\newtheorem{lemma}{Lemma}

\usepackage[norelsize,linesnumbered,ruled]{algorithm2e}
\SetKwRepeat{Do}{do}{while}
\makeatletter
\newcommand{\removelatexerror} {\let\@latex@error\@gobble}
\makeatother

\newcommand{\superscript}[1]{^{\mathrm{#1}}}
\newcommand{\subscript}[1]{_{\mathrm{#1}}}

\usepackage[textsize=tiny,colorinlistoftodos]{todonotes}
\makeatletter
\define@key{todonotes}{bh}[]{
	\setkeys{todonotes}{author=\textbf{Bin}, color=lime!30}}%
\define@key{todonotes}{yz}[]{
	\setkeys{todonotes}{author=\textbf{Yao}, color=blue!30}}%
\makeatother

\newif\ifapp
\appfalse
\apptrue

\newcommand\alice{\emph{Alice}}
\newcommand\bob{\emph{Bob}}
\newcommand\eve{\emph{Eve}}

\begin{document}

\title{Non-Orthogonal Multiplexing in the FBL Regime Enhances Physical Layer Security with Deception}

\author{
	\IEEEauthorblockN{
		Bin~Han\IEEEauthorrefmark{1},
		Yao~Zhu\IEEEauthorrefmark{2},
		Anke~Schmeink\IEEEauthorrefmark{2},
		and~Hans~D.~Schotten\IEEEauthorrefmark{1}\IEEEauthorrefmark{3}
	}
	\IEEEauthorblockA{
		\IEEEauthorrefmark{1}RPTU Kaiserslautern-Landau, \IEEEauthorrefmark{2}RWTH Aachen University, \IEEEauthorrefmark{3}German Research Center of Artificial Intelligence (DFKI)
	}
}

\bstctlcite{IEEEexample:BSTcontrol}

\maketitle

\begin{abstract}
We propose a new security framework for \ac{pls} in the \ac{fbl} regime that incorporates deception technology, allowing for active countermeasures against potential eavesdroppers. Using a symmetric block cipher and power-domain \ac{nom}, our approach is able to achieve high secured reliability while effectively deceiving the eavesdropper, and can benefit from increased transmission power. This work represents a promising direction for future research in \ac{pls} with deception technology.
\end{abstract}


\IEEEpeerreviewmaketitle

\glsresetall

\section{Introduction}\label{sec:introduction}
\Ac{pls} is a rapidly growing field in wireless communications. It aims at securing information transmission by exploiting the characteristics of physical channels, without relying on cryptographic algorithms. Providing a new level of security and privacy, \ac{pls} is becoming increasingly important in today's wireless networks~\cite{HFA2019classification}.

While most research works on \ac{pls} are with the assumption of infinite blocklength codes, recent advance in~\cite{Yang_wiretap_2019} characterizes the \ac{pls} performance for \ac{fbl} codes.   
Based on that, various efforts, such as~\cite{zhu_PLS_2023, Wang_PLS_throughput_2019} have been exploring  in \ac{fbl} regime.
These works have provided insights into the impact of blocklength on \ac{pls} and have shown that \ac{pls} can still be achieved with \ac{fbl}.

Another emerging cluster of research focuses on the application of \ac{noma} in \ac{pls}. \ac{noma} is a promising technology that allows multiple users to share the same frequency and time resources, which can significantly increase spectral efficiency. Especially for \ac{pls}, the interference caused by the superposition signals could be beneficial to improve the security~\cite{Cao_PLS_NOMA_jamming_2021,Xiang_PLS_NOMA_2019}. Therefore, \ac{noma}-based \ac{pls} has been shown to provide enhanced security compared to conventional approaches. Nevertheless, such studies are also generally considering long codes, leaving \ac{noma}-\ac{pls} in the \ac{fbl} regime a virgin land of research. 

Furthermore, the discipline of \ac{pls} has so far been developed as a passive approach to defend against possible eavesdropping, without any capability of detecting or actively countering eavesdroppers. A possible way to make up for this shortcoming is to introduce the deception technologies, which aim to mislead and distract potential eavesdroppers by creating fake data or environments, while keeping the real data and environment secure~\cite{WL2018cyber}. Such technologies can be even deployed to lure eavesdroppers into exposing themselves. However, to the best of our knowledge, there has been so far no reported effort to merge deception technology with \ac{pls}.

In this work, we propose a novel security framework that combines \ac{nom}, \ac{pls} and deception. Using a symmetric block cipher and power-domain multiplexing the ciphered codeword together with the key, we make it possible to deceive eavesdroppers and actively counteract their attempts to intercept transmitted messages. Leveraging the features of \ac{pls} in the \ac{fbl} regime, we can jointly optimize the encryption coding rate and the power allocation, to simultaneously achieve high secured reliability and effective deception.

The remaining part of this paper is organized as follows. We begin with setting up the models and formulating the joint optimization problem in Sec.~\ref{sec:problem}, then analyze the problem to reduce its complexity and propose our solution in Sec.~\ref{sec:approach}. Afterwards, in Sec.~\ref{sec:num_results} we numerically verify our analytical conclusions and evaluate our approach in various aspects of performance, before closing this paper with our conclusion and some outlooks in Sec.~\ref{sec:conclusion}.


\section{Problem Setup}\label{sec:problem}
\subsection{System Model}
We consider a peer-to-peer communication system where information source \alice\ sends messages to the desired receiver \bob\ over a wireless channel $h_\bob$ with gain $z_\bob$, while a potentially existing eavesdropper \eve\ tires to obtain the messages by listening to the side-channel $h_\eve$ with gain $z_\eve$. In this study we consider $z_\bob\geqslant z_\eve>0$, which is a necessary condition of \ac{pls} feasibility and can be generally achieved through appropriate beamforming.

To enable deception, \alice\ encrypts every message with a symmetric block encryption algorithm $f: \mathbb{P}\times\mathbb{K}\overset{f}{\to}\mathbb{P}$, where $\mathbb{P}$ is the set of all possible \emph{payload} messages, and $\mathbb{K}$ the set of all \emph{keys}. Note that every ciphertext is still in the domain of plaintext $\mathbb{P}$. Every message $m\in\mathcal{P}$ is of $d\subscript{M}$ bits, and every key $k\in\mathcal{K}$ of $d\subscript{K}$ bits.  Especially, we assume that $f$ fulfills 
\begin{equation}
	f(p,k)\neq f(p,k'),\quad \forall (p,k,k')\in\mathbb{P}\times\mathbb{K}^2, k\neq k',
\end{equation}
and consider that the sets $\mathbb{P}$, $\mathbb{K}$ and the encrypting algorithm $f$ are known to both \bob\ and \eve.

\begin{figure}[!htbp]
	\centering
	\includegraphics[width=.65\linewidth]{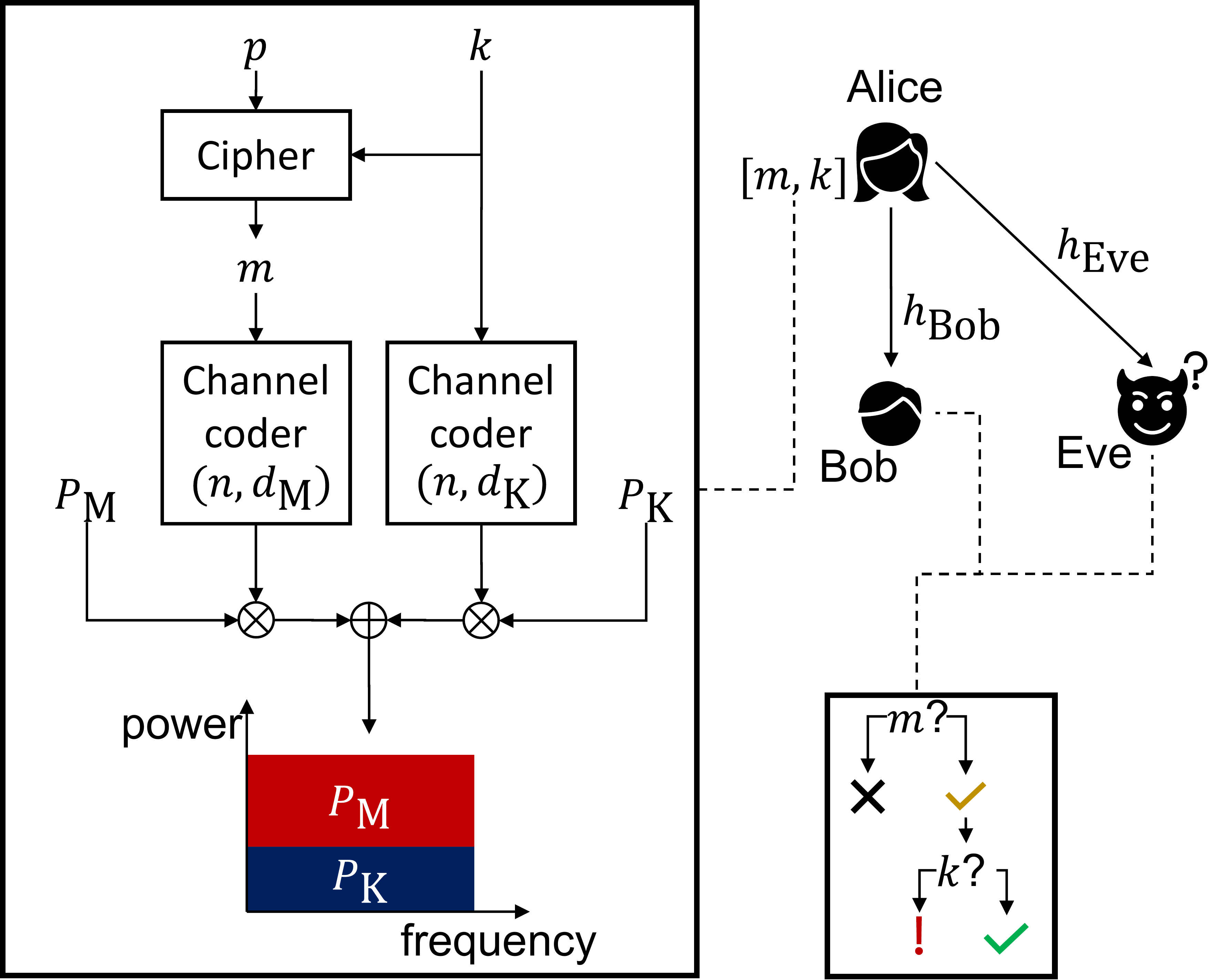}
	\caption{System model}
	\label{fig:model}
\end{figure}

Given a payload $p\in\mathbb{P}$ for \bob, \alice\ randomly selects a key $k$ from $\mathbb{K}$ to cipher it into a message $m=f(p,k)\in\mathbb{P}$. Both the ciphered message $m$ and the key $k$ are then individually encoded by a channel encoder into packets of a finite blocklength $n$. In this study we consider $d\subscript{M}\leqslant n$, $d\subscript{K}\leqslant n$, and $n\geqslant 10$. The two packets are then transmitted together to \bob\ in a power-domain \ac{nom} fashion with $P\subscript{M}>P\subscript{K}$, where $P\subscript{M}$ and $P\subscript{K}$ are the transmission powers for the ciphered message and the key, respectively. Thus, \bob\ (and \eve\ as well) is supposed to carry out \ac{sic} to successively decode $m$ and $k$ under presence of an \ac{awgn} with the power $\sigma^2>0$. When both $m$ and $k$ are successfully decoded, the original payload $p$ can be obtained by $p=f^{-1}(m,k)$, and a unit utility $U=1$ is obtained; when \bob/\eve\ fails to decode $m$, the message is dropped; when \bob/\eve\ successfully decodes $m$ but fails to decode $k$, a false payload will be obtained by deciphering $m$ with an incorrect key $k'\neq k$, so that the receiver (\bob/\eve) is deceived and obtains a unit penalty $U=-1$. The complete transmission and en/decryption procedure is shown in Fig.~\ref{fig:model}.

\subsection{Error and Utility Models}
Considering a finite blocklength $n$  for both packets, we adopt the \emph{Polyanskiy} bound~\cite{Polyanskiy_FBL_2010} to characterize the error rate in \ac{fbl} regime. Given a $n$-symbol codeword of $d$ bits payload, the \ac{per} is upper-bounded by
	$\varepsilon=Q\left[\sqrt{\frac{n}{V}}\left(\eta-\frac{d}{n}\right)\ln 2\right]$,
where $Q(x)=\frac{1}{2}\text{erfc}\left(\frac{x}{\sqrt{2}}\right)$ is the Gaussian tail distribution function, $\gamma$ is the \ac{sinr}, and $V$ is the channel dispersion. For \ac{awgn} channels,
$V(\gamma)=1-\frac{1}{(1+\gamma)^2}$.
$\eta=\frac{\mathcal{C}}{B}$ is the spectral efficiency, where $\mathcal{C}=B\log_2(1+\gamma)$ is the \emph{Shannon} capacity. For \ac{fbl}, we usually normalize the bandwidth to $B=1$ for convenience of analysis, so that
\begin{equation}\label{eq:FBL_err}
	\varepsilon=Q\left[\sqrt{\frac{n}{V(\gamma)}}\left(\mathcal{C}-\frac{d}{n}\right)\ln 2\right].
\end{equation}

For both $i\in\{\bob, \eve\}$, the \ac{sic} begins with decoding the message $m$, where the key $k$ plays the role of interference. Thus, the \ac{sinr} is
$\gamma_{i,\text{M}}=\frac{z_iP\subscript{M}}{z_iP\subscript{K}+\sigma^2}$
and the \ac{per} is
\begin{equation}\label{eq:per_message}
	\epsilon_{i,\text{M}}=\varepsilon_{i,\text{M}}=Q\left[\sqrt{\frac{n}{V(\gamma_{i,\text{M}})}}\left(\mathcal{C}_{i,\text{M}}-\frac{d\subscript{M}}{n}\right)\ln 2\right].
\end{equation}

Upon a successful decoding of $m$, $i$ can carry out the \ac{sic}, and therewith further decode $k$ without being interfered by $m$. In this case, the \ac{snr} is
$\gamma_{i,\text{K}}=\frac{z_iP\subscript{K}}{\sigma^2}$
and the \ac{per} is 
\begin{equation}
	\varepsilon_{i,\text{K}}=Q\left[\sqrt{\frac{n}{V(\gamma_{i,\text{K}})}}\left(\mathcal{C}_{i,\text{K}}-\frac{d\subscript{K}}{n}\right)\ln 2\right].
\end{equation}
Alternatively, in case the decoding of $m$ fails, $i$ can also attempt to directly decode $k$ under the interference from $m$, where the \ac{sinr} is
$\gamma'_{i,\text{K}}=\frac{z_iP\subscript{K}}{z_iP\subscript{M}+\sigma^2}$
and the \ac{per} is
\begin{equation}
	\varepsilon'_{i,\text{K}}=Q\left[\sqrt{\frac{n}{V(\gamma'_{i,\text{K}})}}\left(\mathcal{C}_{i,\text{K}}-\frac{d\subscript{K}}{n}\right)\ln 2\right].
\end{equation}
Thus, the overall error probability in decoding $k$ is
\begin{equation}
	\epsilon_{i,\text{K}} =(1-\varepsilon_{i,\text{M}})\varepsilon_{i,\text{K}}+ \varepsilon_{i,\text{M}}\varepsilon'_{i,\text{K}}.
	\label{eq:per_key}
\end{equation}
As we force $P\subscript{M}>P\subscript{K}$, it always holds that $\gamma'_{i,\text{K}}<\SI{0}{\dB}$.  The direct decoding of $k$ without \ac{sic} is therefore unlikely to succeed due to the strong interference, i.e. we can approximately consider $\varepsilon'_{i,\text{K}}\approx 1$. Moreover, following the classical \ac{fbl} approach we neglect the second-order error term $\varepsilon_{i,\text{M}}\varepsilon_{i,\text{K}}\approx 0$. Applying both approximations on Eq.~\eqref{eq:per_key}, we have
\begin{equation}
	\epsilon_{i,\text{K}}\approx\varepsilon_{i,\text{K}}+ \varepsilon_{i,\text{M}}.
	\label{eq:per_key_approx}
\end{equation}

The expected utility received by both $i\in\{\bob, \eve\}$ is
\begin{equation}\label{eq:expected_utility}
	\begin{split}
		\mathbb{E}\{U_i\}
		=&\left(1-\epsilon_{i,\text{M}}\right)\left(1-\epsilon_{i,\text{K}}\right)-\left(1-\epsilon_{i,\text{M}}\right)\epsilon_{i,\text{K}}\\
		=&\left(1-\epsilon_{i,\text{M}}\right)\left(1-2\epsilon_{i,\text{K}}\right),
	\end{split}
\end{equation}
and we consider the system's overall utility
$U_\Sigma = U_\bob-U_\eve$.

\subsection{Strategy Optimization}
Now consider a fixed power budget $P_\Sigma\in(0,+\infty)$ of \alice, a fixed packet size $n$,
and a fixed payload message length  $d\subscript{M}$. We look for an optimal strategy of encryption coding and power allocation that maximizes the system utility:
\begin{maxi!}[2]
	{d\subscript{K}, P\subscript{M}, P\subscript{K} }{\mathbb{E}\{U_\Sigma\}}{\label{prob:max_utility}}{}
	\addConstraint{P\subscript{M}\geqslant 0 \label{con:non-neg_power_message}}
	\addConstraint{P\subscript{K}\geqslant 0 \label{con:non-neg_power_key}}
	\addConstraint{P\subscript{M}+P\subscript{K}\leqslant P_\Sigma\label{con:max_power}}
	\addConstraint{d\subscript{K}\in\{0,1,\dots n\}\label{con:max_key_length}}
	\addConstraint{\epsilon_{\bob,\text{M}}\leqslant \epsilon\superscript{th}_{\bob,\text{M}}\label{con:err_bob_message_threshold}}
	\addConstraint{\epsilon_{\eve,\text{M}}\leqslant \epsilon\superscript{th}_{\eve,\text{M}}\label{con:err_eve_message_threshold}}
	\addConstraint{\epsilon_{\bob,\text{K}}\leqslant \epsilon\superscript{th}_{\bob,\text{K}}\label{con:err_bob_key_threshold}}
	\addConstraint{\epsilon_{\eve,\text{K}}\geqslant \epsilon\superscript{th}_{\eve,\text{K}}\label{con:err_eve_key_threshold}},
\end{maxi!}
where $\epsilon\superscript{th}_{\bob,\text{M}}$, $\epsilon\superscript{th}_{\eve,\text{M}}$, $\epsilon\superscript{th}_{\bob,\text{K}}$, and $\epsilon\superscript{th}_{\eve,\text{K}}$ are pre-fixed thresholds of error probability.

\section{Proposed Approach}\label{sec:approach}
While the multivariate program \eqref{prob:max_utility} is hard to tract, we can derive the following lemma and theorems to reduce its complexity.
\ifapp
	The detailed proofs are provided in the appendices.  
\fi

\begin{theorem}\label{th:full_power_transmission}
	With $\epsilon_{\bob,\text{M}}< 0.5$, $\epsilon_{\eve,\text{M}}< 0.5$, and $P_\Sigma<+\infty$, given any $d\subscript{K}$, the optimal power allocation $P\superscript{o}\subscript{M}$ and $P\superscript{o}\subscript{K}$ must fulfill $P\superscript{o}\subscript{M}+P\superscript{o}\subscript{K}=P_\Sigma$. 
\end{theorem}

Driven by Theorem~\ref{th:full_power_transmission}, we define the expected $U_\Sigma$ under full-power transmission%
~as $U\subscript{FP}\triangleq\mathbb{E}\{U_\Sigma\vert {P\subscript{K}=P_\Sigma - P\subscript{M}}\}$,
and Problem~\eqref{prob:max_utility} is degraded to bivariate:
\begin{maxi!}[2]
	{d\subscript{M}, P\subscript{M}}{U\subscript{FP}}{\label{prob:max_utility_degraded}}{}
	\addConstraint{P\subscript{K}\in[0,P_\Sigma]\label{con:message_power}}
	\addConstraint{\text{constraints \eqref{con:err_bob_message_threshold}--\eqref{con:err_eve_key_threshold}}\nonumber}.
\end{maxi!}
However, Problem~\eqref{prob:max_utility_degraded} is still a mixed integer non-convex problem, which is difficult to solve. To tackle this issue, we relax $d\subscript{K}$ from integer into a real value, i.e., $0\leqslant d\subscript{K}\leqslant n$. Then, we leverage the \ac{bcd} framework to obtain the corresponding solutions iteratively. 

In particular, in each $t\superscript{th}$ iteration, we fix the bit length of key as $d\subscript{K}=d^{(t-1)}\subscript{K}$. Then, Problem~\eqref{prob:max_utility_degraded} is reformulated as
\begin{maxi!}[2]
	{P\subscript{M}}{U\subscript{FP}}{\label{prob:max_utility_degraded_fix_d}}{}
	\addConstraint{P\subscript{M}\in[0,P_\Sigma]}
    \addConstraint{d\subscript{K}=d^{(t-1)}\subscript{K}}
	\addConstraint{\text{constraints \eqref{con:err_bob_message_threshold}--\eqref{con:err_eve_key_threshold}}\nonumber}.
\end{maxi!}
Note that $U\subscript{FP}$ consists of the multiplications and subtraction of \glspl{per} for both $m$ and $k$. Thus, we first characterize their convexity and monotonicity with the following Lemma: 
\begin{lemma}\label{lemma:convexity_power}
	With $\epsilon_{\bob,\text{M}}< 0.5$, $\epsilon_{\eve,\text{M}}< 0.5$, and $P_\Sigma<+\infty$, both $\epsilon_{\bob,\text{M}}$ and $\epsilon_{\eve,\text{M}}$ are strictly monotonically decreasing and convex of $P\subscript{M}$ in the feasible region of Problem~\eqref{prob:max_utility_degraded}.
 \end{lemma}
Therewith, the constraints~\eqref{con:err_bob_message_threshold}--\eqref{con:err_eve_key_threshold} are convex while the rest of them being affine. Furthermore, we can establish the following partial concavity of the utility $U\subscript{FP}$: 
\begin{theorem}\label{th:concavity_message_power}
	$U\subscript{FP}$ is concave of $P\subscript{M}$ in the feasible region of Problem~\eqref{prob:max_utility_degraded}.
\end{theorem}
As a result, Problem~\eqref{prob:max_utility_degraded_fix_d} is concave and can be solved efficiently with any standard convex optimization tool. Denoting its optimum $P^{(t)}\subscript{M}$, we fix $P\subscript{M}=P^{(t)}\subscript{M}$ in Problem~\eqref{prob:max_utility_degraded} as:
\begin{maxi!}[2]
	{d\subscript{K}}{U\subscript{FP}}{\label{prob:max_utility_degraded_fix_P}}{}
    \addConstraint{d\subscript{K}\in[0,n]}    \addConstraint{P\subscript{M}=P^{(t)}\subscript{M}}
	\addConstraint{\text{constraints \eqref{con:err_bob_message_threshold}--\eqref{con:err_eve_key_threshold}}\nonumber}.
\end{maxi!}
We can also identify the following partial concavity of $U\subscript{FP}$:
\begin{theorem}\label{th:concavity_key_length}
	If $\epsilon\superscript{th}_{\bob,\text{K}}\leqslant 0.5$ and $\epsilon\superscript{th}_{\eve,\text{K}}\geqslant 0.5$, $U\subscript{FP}$ is concave of $d\subscript{K}$ in the feasible region of Problem~\eqref{prob:max_utility_degraded}.
\end{theorem}
Accordingly, Problem~\eqref{prob:max_utility_degraded_fix_P} can be solved as a concave problem, since the objective function is concave and the constraints are convex or affine. We denote the optimal solution of Problem~\eqref{prob:max_utility_degraded_fix_P} as $d^{(t)}\subscript{K}$, which is set as the fixed value of $d\subscript{K}$ is the $(t+1)\superscript{th}$ iteration. Moreover, its corresponding optimal value is denoted as $U^{(t)}\subscript{LF}=U_{LF}(P^{(t)}\subscript{M},d^{(t)}\subscript{K})$. This process will repeat until it meets either the stop criterion $\left\vert U^{(t)}-U^{(t-1)}\right\vert\leqslant \mu$ or a given maximal allowed iteration rounds $T$, where $\mu$ is a non-negative threshold. The obtained solutions are denoted as $P^{*}\subscript{K}$ and $d^{*}\subscript{K,R}$, respectively. Specially, we initialize the variable pair as $\left(d^{(0)}\subscript{K},P^{(0)}\subscript{M}\right)= \left(d\superscript{init}\subscript{K},P\superscript{init}\subscript{M}\right)$ and the obtained utility $U^{(0)}\subscript{FP}=-\infty$. It should be emphasized that the initial value $(d\superscript{init}\subscript{K},P\superscript{init}\subscript{M})$ must be feasible for Problem~\eqref{prob:max_utility_degraded}.  
Recalling that $d\subscript{K}$ must be integer, the optimal integer solution shall be obtained via comparing the integer neighbors of  $d^{*}\subscript{K,R}$:
\begin{equation}
    d^*\subscript{K}=\arg\max\limits_{m\in\left\{\left\lfloor d^{*}\subscript{K,R}\right\rfloor,\left\lceil d^{*}\subscript{K,R}\right\rceil\right\} } U\subscript{LF}(P^{*}\subscript{M}).
\end{equation}
The \ac{bcd} framework to solve Problem~\eqref{prob:max_utility_degraded} can be described by Algorithm~\ref{alg:bcd}. It is able to achieve sub-optimal solutions with the complexity of $\mathcal{O}(\phi(4N^2))$, where $N$ is the number of variables in Problem~\eqref{prob:max_utility_degraded} and $\phi(\cdot)$ represents the iteration numbers upon the solution accuracy~\cite{Tseng_BCD_2001}.  
\begin{algorithm}[!htpb]
	\caption{The \ac{bcd} framework}
	\label{alg:bcd}
	\scriptsize
	\DontPrintSemicolon
	Input: $\mu, T, P_\Sigma, d\subscript{M}, n$\; 
	Initialize: $t=1, P\superscript{0}\subscript{M}=P\superscript{init}\subscript{M}, d\superscript{0}\subscript{K}=d\superscript{init}\subscript{K}, U\superscript{o}\subscript{FP}=-\infty$\;
	\Do(\tcp*[f]{Check convergence}){$|U^{(t)}-U^{(t-1)}|\geqslant\xi$}{
		\uIf(\tcp*[f]{Limiting the number of iterations}){$t\leqslant T$}{
			$P^{(t)}\subscript{M} \gets \arg\max\limits_{P\subscript{M}}U\subscript{FP}\left(d^{(t-1)}\subscript{K}, P\subscript{M}\right)$
			$d^{(t)}\subscript{K} \gets \arg\max\limits_{d\subscript{K}}U\subscript{FP}\left(d\subscript{K}, P^{(t)}\subscript{M}\right)$\;
    			$U^{(t)}\subscript{FP}\gets U\subscript{FP}\left(d^{(t)}\subscript{K}, P^{(t)}\subscript{M}\right)$\;
			$t\gets t+1$\;
		}
		\Else{\textbf{break}\;}
	}
        $P^*\subscript{K}\gets P^{(t)}\subscript{K}$
        
        $d^*\subscript{K,R}\gets \arg\max\limits_{m\in\left\{\left\lfloor d^{(t)}\subscript{K} \right\rfloor,\left\lceil d^{(t)}\subscript{K}\right\rceil\right\} } U\subscript{LF}(P^{(t)}\subscript{M})$
        
\Return{$\left(d^*\subscript{K}, P^*\subscript{M}\right)$}
\end{algorithm}

\section{Numerical Verification}\label{sec:num_results}
To verify our analyses and evaluate our proposed approach, we conducted a series of numerical simulations. All these simulations share the same setup listed in Tab.~\ref{tab:sim_setup}. Task-specific configurations will be provided correspondingly later when we introduce each of them in details below.
\begin{table}
	\centering
	\caption{Simulation setup}
	\label{tab:sim_setup}
	\begin{tabular}{m{1.7cm} l l}
		\toprule[2px]
		\textbf{Parameter} 		&\textbf{Value} 		&\textbf{Remark}\\
		\midrule[1px]
		$\sigma^2$					&\SI{1}{\milli\watt}			&\Ac{awgn} power\\
		$z_\bob$					  &\SI{0}{\dB}				&Channel gain of \bob\\
		$B$								 &\SI{1}{\hertz}			&Normalized to unity bandwidth\\
		$n$								 &64							 &Block length per packet\\
		$\epsilon_{\bob,\text{M}}\superscript{th}$, $\epsilon_{\bob,\text{K}}\superscript{th}$, $\epsilon_{\eve,\text{K}}\superscript{th}$, $\epsilon_{\eve,\text{K}}\superscript{th}$		&0.5							 &Threshold in constraints \eqref{con:err_bob_message_threshold}--\eqref{con:err_eve_key_threshold}\\
		$\xi$							&$2\times 10^{-16}$	&\ac{bcd} convergence threshold\\
		$K$								&100							&Maximal number of iterations in \ac{bcd}\\
		\bottomrule[2px]
	\end{tabular}
\end{table}

\subsection{Superiority of Full-Power Transmission}
\begin{figure}[!htpb]
	\centering
	\includegraphics[width=\linewidth]{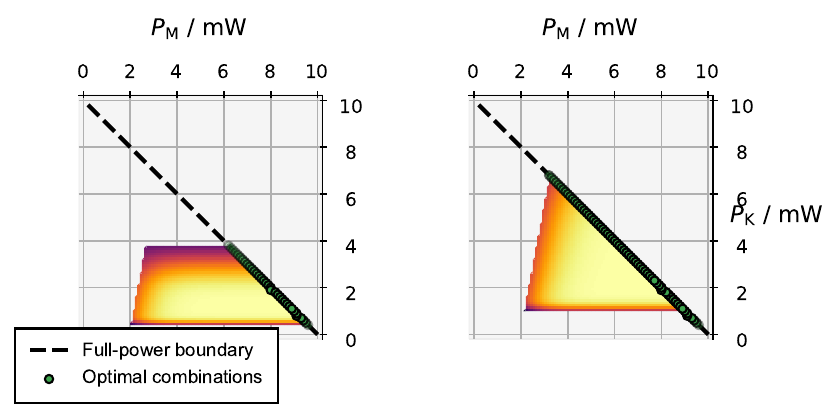}
	\caption{The optimal selection of $P\subscript{M}$ upon different specifications of $P\subscript{K}$, with $d\subscript{K}=30$ (left) and $d\subscript{K}=60$ (right).}
	\label{fig:full_power_demo}
\end{figure}

First, to verify Theorem~\ref{th:full_power_transmission} that the optimal power allocation always fully exploits the power budget $P_\Sigma$, we set $P_\Sigma=\SI{10}{\milli\watt}$, $z_\eve=\SI{-10}{\dB}$, $d\subscript{M}=16$, and computed $\mathbb{E}\{U_\Sigma\}$ w.r.t. Eq.~\eqref{eq:expected_utility} in the region $(P\subscript{M}, P\subscript{K})\in[0,\SI{10}{\milli\watt}]^2$. For each $P\subscript{M}$ in the feasible region of Problem \eqref{prob:max_utility}, we executed exhaustive search to find the optimal $P\superscript{o}\subscript{K}$ that maximizes $\mathbb{E}\{U_\Sigma\}$. We carried out this test twice, with different settings of the ciphering key length $d\subscript{K}$: once with 30 bits and once with 60 bits. As the results illustrated in Fig.~\ref{fig:full_power_demo} reveal, in both cases, all optimal power allocations land on the full-power boundary $P\subscript{M}+P\subscript{K}=P_\Sigma$, which supports our analysis.

\subsection{Utility Surface}
\begin{figure}[!htpb]
	\centering
	\includegraphics[width=.8\linewidth]{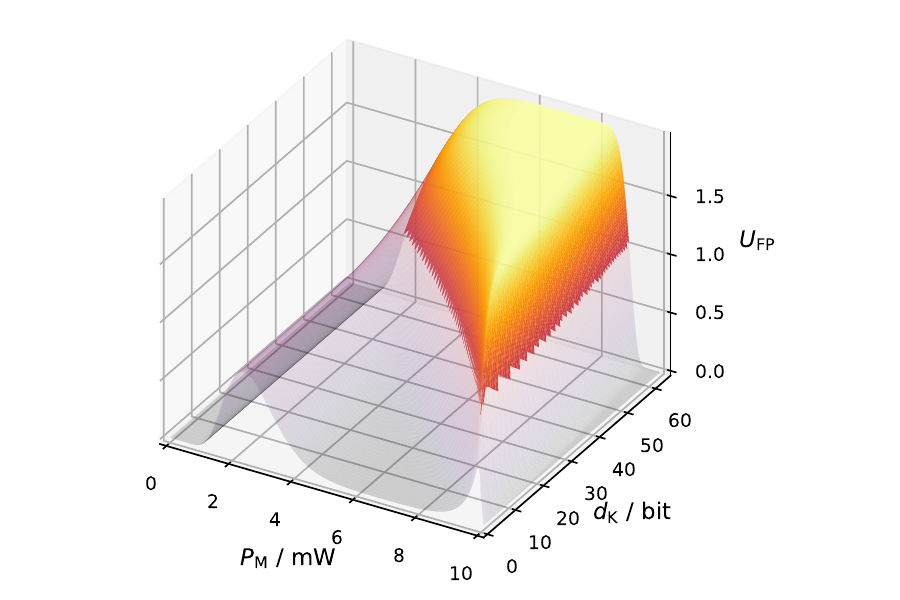}
	\caption{The system utility $U\subscript{FP}$ under full-power transmission}
	\label{fig:concavity}
\end{figure}

To obtain insights about the overall surface of system utility $U\subscript{FP}$ under the strategy of full-power transmission, we set $P_\Sigma=\SI{10}{\milli\watt}$, $z_\eve=\SI{-10}{\dB}$, $d\subscript{M}=16$, and computed $U\subscript{FP}$ in the region $(P\subscript{M}, d\subscript{K})\in[0,\SI{10}{\milli\watt}]\times\{0,1,\dots 64\}$. The result is depicted in Fig.~\ref{fig:concavity}, where the feasible region outlined by \eqref{con:err_eve_message_threshold}--\eqref{con:err_eve_key_threshold} is highlighted with higher opacity w.r.t. the rest parts. We can observe from the figure that $U\subscript{FP}$ is concave of both $P\subscript{M}$ and $d\subscript{K}$ within the feasible region, while the convexity/concavity outside the region is rather complex.

\subsection{Convergence Test of the BCD Framework}
\begin{figure}[!htpb]
	\centering
	\includegraphics[width=.9\linewidth]{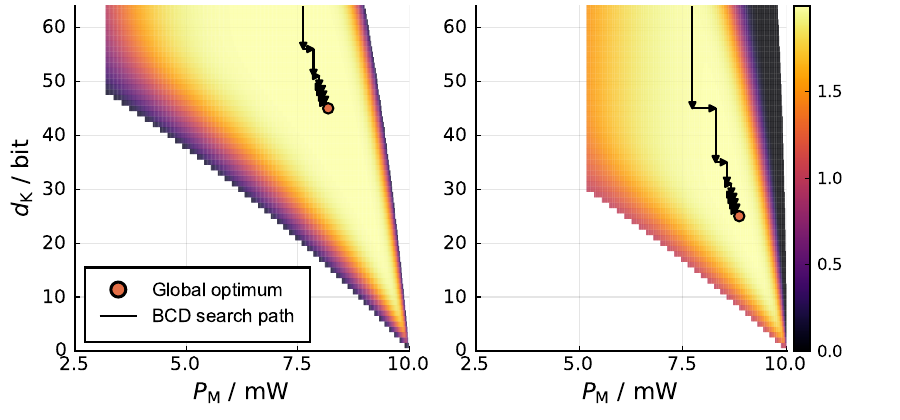}
	\caption{The $U\subscript{FP}$ surface and the search path of Algorithm~\ref{alg:bcd}, with $d\subscript{M}=16$ (left) and $d\subscript{M}=24$ (right).}
	\label{fig:bcd_demo}
\end{figure}

To monitor the feasibility of the proposed \ac{bcd} framework in the joint optimization of the encryption coding and power allocation, we set $P_\Sigma=\SI{10}{\milli\watt}$, $z_\eve=\SI{-10}{\dB}$, and tested our Algorithm~\ref{alg:bcd} with two sample configurations of the payload message bit length: $d\subscript{M}=16$ and $d\subscript{M}=24$. As the results shown in Fig.~\ref{fig:bcd_demo} are suggesting, the \ac{bcd} algorithm efficiently converges in both cases, and successfully achieves the global optima after 8 and 9 iterations, respectively.

\subsection{Performance Evaluation}
To assess the security level and deceiving capability of our proposed approach, we are interested in two different performance indicators of it. Regarding secured transmission, we consider the secure reliability 
\begin{equation}
	R\subscript{s}=(1-\epsilon_\bob)\epsilon_\eve,
\end{equation}
where $\epsilon_i=1-(1-\epsilon_{i,\text{M}})(1-\epsilon_{i,\text{K}})$ is the overall error rate in decoding the payload message for both $i\in\{\bob, \eve\}$. Regarding deception, we consider the effective deception rate
\begin{equation}
	R\subscript{d}=(1-\delta_\bob)\delta_\eve,
\end{equation}
where $\delta_i=(1-\epsilon_{i,\text{M}})\epsilon_{i,\text{K}}$ is the probability that $i\in\{\bob, \eve\}$ is deceived to obtain an incorrect payload.

\begin{figure}[!htpb]
	\centering
	\includegraphics[width=.8\linewidth]{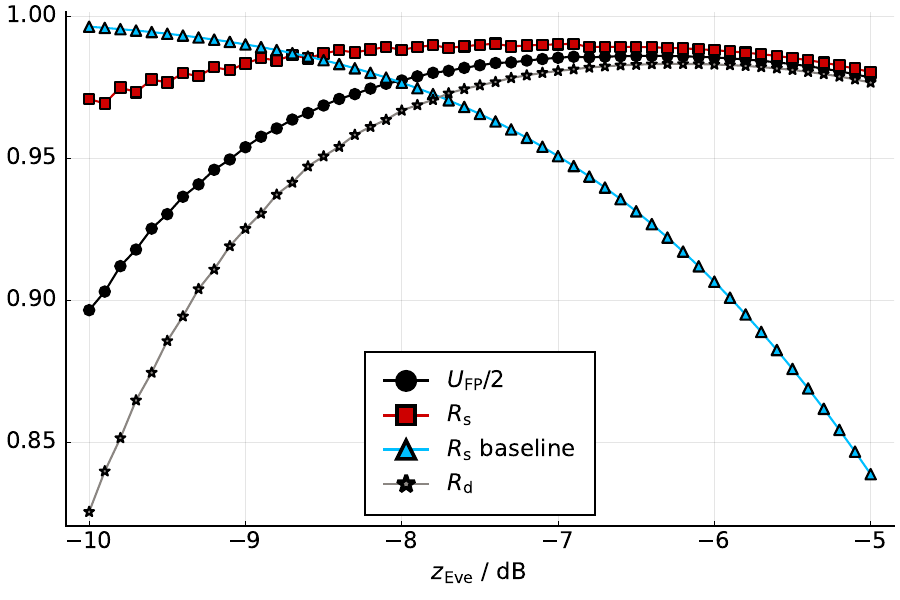}
	\caption{Results of sensitivity test regarding $z_\eve$}
	\label{fig:numerical_performance_h_eve}
\end{figure}

First we set $P_\Sigma=\SI{3}{\milli\watt}$, $d\subscript{M}=16$, and measured $U\subscript{FP}$, $R\subscript{s}$, as well as $R\subscript{d}$ under different conditions of $z_\eve$. As a baseline to compare with, we also implemented a classical secure-reliability-optimal \ac{pls} solution without the deceiving mechanism, i.e., we set $d\subscript{K}=0$, $P\subscript{K}=0$, and chose the optimal $P\subscript{M}\in[0,P_\Sigma]$ that maximizes $R\subscript{s}$. The results are shown in Fig.~\ref{fig:numerical_performance_h_eve}. Compared to the baseline solution, our approach offers a significant enhancement in the robustness of $R_s$ against increasing $z_\eve$. When \eve\ has a poor channel with \SI{-10}{\dB} gain, our solution compromises only slightly regarding the secure reliability, by less than $3\%$ compared to the baseline, while delivering a high effective deception rate over $80\%$. Moreover, the deception rate may even benefit from a good channel condition of \eve\ (since \eve\ is in this case more likely to decode the message packet), exceeding $98\%$ around $z_\eve=\SI{-7}{\dB}$.

\begin{figure}[!htpb]
	\centering
	\includegraphics[width=.8\linewidth]{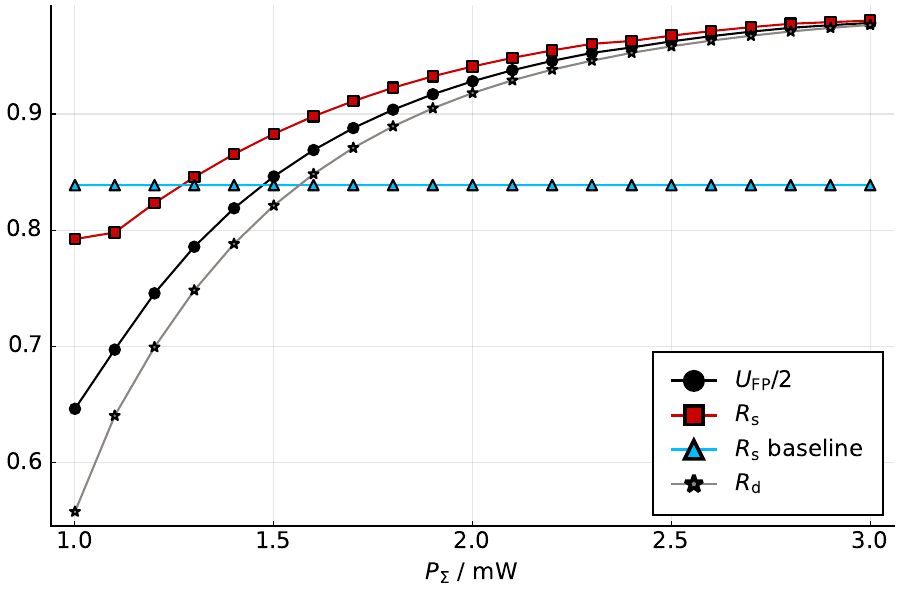}
	\caption{Results of sensitivity test regarding $P_\Sigma$}
	\label{fig:numerical_performance_p_total}
\end{figure}

Then we fixed $z_\eve=\SI{-5}{\dB}$ and repeated the assessment under different power budgets $P_\Sigma$, whereby we obtained the results in Fig.~\ref{fig:numerical_performance_p_total}. Again, with sufficient power budget, we observe a significant enhancement in secure reliability w.r.t. the baseline, in addition to a high effective deception rate. Moreover, in contrast to the classical \ac{pls} solutions that cannot benefit from higher power budget, our approach can be overall enhanced in various aspects of performance by raising $P_\Sigma$.

\begin{figure}[!htpb]
	\centering
	\includegraphics[width=.9\linewidth]{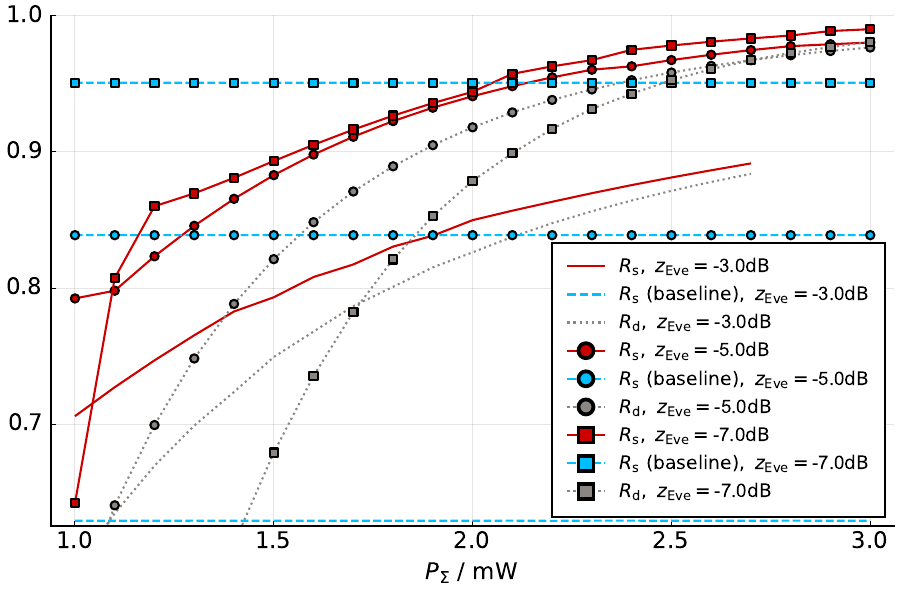}
	\caption{Benchmark results}
	\label{fig:benchmark}
\end{figure}

Results of a more comprehensive benchmark test, which mixes different settings of $z_\eve$ and $P_\Sigma$, are illustrated in Fig.~\ref{fig:benchmark}. It reveals that our approach generally outperforms the baseline of classical \ac{pls} regarding the secured reliability, as long as supported by a sufficient power budget. More specifically, the minimal $P_\Sigma$ required for our solution to outperform
the baseline increases along with the channel gain gap $z_\bob-z_\eve$. In addition, it is worth to discuss the case when $z_\eve=\SI{-3}{\dB}$ and $P_\Sigma\geqslant \SI{2.7}{\milli\watt}$, where no feasible solution with full-power transmission can be found under the constraints \eqref{con:max_key_length} -- \eqref{con:err_eve_key_threshold}. In such cases, we suggest to either take a sub-optimal solution with lower transmission power that $P\subscript{M}+P\subscript{K}<P_\Sigma$, or adjust the blocklength $n$ of each packet.

\section{Conclusion and Outlooks}\label{sec:conclusion}
In this paper, we have proposed a novel security framework to enhance the classical \ac{pls} approach with the capability of deceiving eavesdroppers, and provided a solution to jointly optimize its encryption coding rate and power allocation. With numerical results we have demonstrated the effectiveness of our methods. Compared to conventional \ac{pls} approaches, our proposal exhibits a unique feature of benefiting from higher transmission power budget, which makes it superior when confronting eavesdroppers with good channels. 

As a pioneering research, this work can be extended in multiple directions. First, the migration of our proposed framework to orthogonal multiplexing is of great interest. Second, the issue that the feasible region vanishes under good eavesdropping channel and high power budget shall be addressed. Third, a more generic and flexible solution can be achieved by modifying the objective and constraints of the optimization problem. Furthermore, specific design of the encryption codec must be investigated to efficiently realize our approach.

\section*{Acknowledgment}
This work is supported in part by the German Federal Ministry of Education and Research in the programme of “Souverän. Digital. Vernetzt.” joint projects 6G-RIC (16KISK028), Open6GHub (16KISK003K/16KISK004/16KISK012), in part by the German Research Council through the basic research project under grant number DFG SCHM 2643/17, and in part by the European Commission via the Horizon Europe project Hexa-X-II (101095759). Y. Zhu (yao.zhu@rwth-aachen.de) is the corresponding author. 

\ifapp
	\appendices
	\section{Proof of Theorem~\ref{th:full_power_transmission}}
	\begin{proof}
		This theorem can be proven straightforwardly by the contradiction. Given a certain $d\subscript{K}$, in this appendix we can use the following notations for convenience:
		\begin{align}
			&\overline{U}_\Sigma(P_1,P_2)\triangleq\mathbb{E}\left\{U_\Sigma\vert P\subscript{M}=P_1, P\subscript{K}=P_2\right\},\\
			&\epsilon_{i,j}(P)\triangleq\epsilon_{i,j}\vert_{P_j=P}, \quad\forall (i,j)\in\{\bob,\eve\}\times\{\text{M},\text{K}\}.
		\end{align}
		Suppose there exists an optimal power allocation $\left(P\superscript{o}\subscript{K}, P\superscript{o}\subscript{M}\right)$ that leaves from the power budget a positive residual $P_\Delta=P_\Sigma-P\superscript{o}\subscript{K}-P\superscript{o}\subscript{M}>0$. Since it is optimal, for all feasible $\left(P'\subscript{M}, P'\subscript{K}\right)\neq \left(P\superscript{o}\subscript{K}, P\superscript{o}\subscript{M}\right)$ it must hold that
		\begin{equation}
			\overline{U}_\Sigma\left(P\superscript{o}\subscript{M}, P\superscript{o}\subscript{K}\right)\geqslant \overline{U}_\Sigma\left(P'\subscript{M}, P'\subscript{K}\right).
		\end{equation}
		Meanwhile, there is always another feasible allocation $(P\superscript{f}\subscript{M}, P\superscript{o}\subscript{K})$ where $P\superscript{f}\subscript{M}=P\superscript{o}\subscript{M}+P_\Delta$. Given the same $P\subscript{K}$, it is trivial to see that $\epsilon_{\bob,\text{M}}$ and $\epsilon_{\eve,\text{M}}$ are monotonically decreasing
		in $P\subscript{M}$, so it always holds that 
		\begin{align}
			\epsilon_{\bob,\text{M}}(P\superscript{o}\subscript{M})&>\epsilon_{\bob,\text{M}}\left(P\superscript{f}\subscript{M}\right),\\
			\epsilon_{\eve,\text{M}}(P\superscript{o}\subscript{M})&>\epsilon_{\eve,\text{M}}\left(P\superscript{f}\subscript{M}\right)
		\end{align}
	 	 Therefore, we have:
		\begin{equation}
			\begin{split}
					&\overline{U}_\Sigma\left(P\superscript{f}\subscript{M}, P\superscript{o}\subscript{K}\right)-\overline{U}_\Sigma\left(P\superscript{o}\subscript{M}, P\superscript{o}\subscript{K}\right)\\
				=&\left[1-2\epsilon_{\bob,\text{K}}\left(P\superscript{o}\subscript{K}\right)\right]\left[\epsilon_{\bob,\text{M}}\left(P\superscript{o}\subscript{M}\right)-\epsilon_{\bob,\text{M}}\left(P\superscript{f}\subscript{M}\right)\right]\\
				&-\left[1-2\epsilon_{\eve,\text{K}}\left(P\superscript{o}\subscript{K}\right)\right]\left[\epsilon_{\eve,\text{M}}\left(P\superscript{o}\subscript{M}\right)-\epsilon_{\eve,\text{M}}\left(P\superscript{f}\subscript{M}\right)\right]\\
				>&0.
			\end{split}
		\end{equation}
		The inequality above holds, since $2\epsilon_{\bob,\text{K}}\leqslant  2\epsilon\superscript{th}_{\bob,\text{K}}< 1$ and $2\epsilon_{\eve,\text{K}}\geqslant  2\epsilon\superscript{th}_{\eve,\text{K}}> 1$. In other words, the solution $P\superscript{f}\subscript{K}$ and $P\superscript{f}\subscript{M}$ achieves a better utility $\overline{U}_\Sigma\left(P\superscript{f}\subscript{M},P\superscript{o}\subscript{K}\right)$ than $\overline{U}_\Sigma\left(P\superscript{o}\subscript{M},P\superscript{o}\subscript{K}\right)$, which violates the assumption of optimum.
	\end{proof}
	
	\section{Proof of Lemma~\ref{lemma:convexity_power}}\label{app:proof_convexity_power}
	\begin{proof}
		For the sake of clarity, we define an auxiliary function $\omega(\gamma)\triangleq\sqrt{\frac{n}{V(\gamma)}}\left[\mathcal{C}(\gamma)-\frac{d}{n}\right]$. In~\cite{Zhu_convex_2022,zhu_PLS_2023}, we have shown that  $\omega$ is monotonically decreasing and convex of the S(I)NR $\gamma$, i.e.,
		\begin{equation}\label{eq:monoton_omega_in_gamma}
			\begin{split}
				&\frac{\partial{\omega}}{\partial{\gamma}}\\=&\sqrt{\frac{n}{V}}\left[\frac{\gamma^2+2\gamma-\ln(\gamma+1)}{(\gamma^2+2\gamma)(\gamma+1)}
				+\frac{d}{2V}\frac{2}{(1+\gamma)^3}\right]\geqslant 0,
			\end{split}
		\end{equation}
		\begin{equation}
			\frac{\partial^{2} \omega}{\partial \gamma^{2}}\leqslant\sqrt{\frac{n}{[\gamma(\gamma+2)]^{5}}}
			\left[0-3 \ln{2}\cdot(\gamma+1) \frac{d}{n}\right] \leqslant 0.
		\end{equation}
		
		Moreover, it is clear that the \ac{fbl} error probability itself, i.e., $\varepsilon$ in Eq.~\eqref{eq:FBL_err}, is decreasing and convex of $\omega$ if $\mathcal{C}-\frac{d}{n}\geqslant 0$, otherwise it is decreasing and concave if $\mathcal{C}-\frac{d}{n}\leqslant 0$.
		This can be verified straightforwardly with $\frac{\partial\varepsilon}{\partial \omega}=-\frac{1}{\sqrt{2\pi}}e^{-\frac{\omega^2}{2}}\leqslant 0 $ and $\frac{\partial^2 \varepsilon}{\partial \omega^2}$. 
		Since $\epsilon_{i,\text{M}}=\varepsilon_{i,\text{M}}$ as given by Eq.~\eqref{eq:per_message}, 	all aforementioned conclusions regarding $\varepsilon$ also hold for $\epsilon_{i,\text{M}}$ with both $i\in\{\bob,\eve\}$. With $P\subscript{M}+P\subscript{K}=P_\Sigma$, we have
		\begin{equation}
			\begin{split}
				&\frac{\partial\gamma_{i,\text{M}}}{\partial P\subscript{M}}
				=\frac{\partial}{\partial P\subscript{M}}\left(\frac{z_iP\subscript{M}}{z_i(P_\Sigma-P\subscript{M})+\sigma^2}\right)\\
				=&\frac{z_i(z_iP_\Sigma+\sigma^2)}{(z_iP\subscript{K}+\sigma^2)^2}>0,
			\end{split}
		\end{equation}
		\begin{equation}
			\frac{\partial^2\gamma_{i,\text{M}}}{\partial P^2\subscript{M}}=\frac{2z_i^2(z_iP_\Sigma+\sigma^2)}{(z_iP\subscript{K}+\sigma^2)^3}>0,
		\end{equation}
		\begin{equation}
			\frac{\partial\epsilon_{i,\text{M}}}{\partial P\subscript{M}}=\frac{\partial\epsilon_{i,\text{M}}}{\partial \omega_{i,\text{M}}}\frac{\partial\omega_{i,\text{M}}}{\partial\gamma_{i,\text{M}}}\frac{\partial\gamma_{i,\text{M}}}{\partial P\subscript{M}}\leqslant0\label{eq:monoton_per_message_power}
		\end{equation}
		Especially, the equity in Eq.~\eqref{eq:monoton_per_message_power} is achieved only when $\omega_{i,\text{M}}\to+\infty$ and $\gamma_{i,\text{M}}\to+\infty$, i.e. $P\subscript{M}\to\infty$. So given a limited $0<P_\Sigma<+\infty$, it always holds $\frac{\partial\epsilon_{i,\text{M}}}{\partial P\subscript{M}}<0$, i.e. $\epsilon_{i,\text{M}}$ is \emph{strictly} monotonically decreasing of $P\subscript{M}$.
		
		Moreover, since $n\geqslant 10$ and $\epsilon_{i,\text{M}}\leqslant \epsilon\superscript{th}_{i,\text{M}}<0.5$, there is 
		\begin{strip}
			\begin{equation}
				\begin{split}
					&\frac{\partial^2 \epsilon_{i,\text{M}}}{\partial P\subscript{M}^2}
					=
					\frac{
						\partial \epsilon_{i,\text{M}}
					}
					{
						\partial \gamma_{i,\text{M}}
					}
					\frac{
						\partial^2 \gamma_{i,\text{M}}
					}
					{
						\partial P\subscript{M}^2
					}
					+\frac{
						\partial^2 \epsilon_{i,\text{M}}
					}
					{
						\partial \gamma^2_{i,\text{M}}
					}
					\left(
					\frac{
						\partial \gamma_{i,\text{M}}
					}
					{
						\partial P\subscript{M}
					}
					\right)^2\\
					=&\frac{1}{\sqrt{2\pi}}e^{-\frac{\omega_{i,\text{M}}}{2}}
					\left(
					\frac{\partial \omega_{i,\text{M}}}{\partial \gamma_{i,\text{M}}}
					\left(
					\omega_{i,\text{M}}
					\frac{\partial \omega_{i,\text{M}}}{\partial \gamma_{i,\text{M}}}
					\left(
					\frac{\partial \gamma_{i,\text{M}}}{\partial P\subscript{M}}        
					\right)^2
					-\frac{\partial^2 \gamma_{i,\text{M}}}{\partial P\subscript{M}^2}
					\right)
					-\underbrace{
						\frac{\partial^2 \omega_{i,\text{M}}}{\partial \gamma^2_{i,\text{M}}}
						\left(
						\frac{\partial \gamma_{i,\text{M}}}{\partial P\subscript{M}}
						\right)^2
					}_{\leqslant 0}
					\right)\\
					\geqslant&\frac{1}{\sqrt{2\pi}}e^{-\frac{\omega_{i,\text{M}}}{2}}
					\frac{\partial \omega_{i,\text{M}}}{\partial \gamma_{i,\text{M}}}
					\left(
					\omega_{i,\text{M}}
					\frac{\partial \omega_{i,\text{M}}}{\partial \gamma_{i,\text{M}}}
					\left(
					\frac{\partial \gamma_{i,\text{M}}}{\partial P\subscript{M}}        
					\right)^2
					-\frac{\partial^2 \gamma_{i,\text{M}}}{\partial P\subscript{M}^2}
					\right)\\
					=&\frac{1}{\sqrt{2\pi}}e^{-\frac{\omega_{i,\text{M}}}{2}}
					\frac{\partial \omega_{i,\text{M}}}{\partial \gamma_{i,\text{M}}}
					\left(
					\omega_{i,\text{M}}
					\frac{\partial \omega_{i,\text{M}}}{\partial \gamma_{i,\text{M}}}
					\frac{
						z^2_1(z_iP_\Sigma+\sigma^2_i)^2
					}{
						(z_iP\subscript{K}+\sigma^2_i)^4
					}
					-\frac{
						2z_i^2(z_iP_\Sigma+\sigma^2_i)
					}{
						(z_iP\subscript{K}+\sigma^2_i)^3
					}
					\right)\\
					=&\frac{1}{\sqrt{2\pi}}e^{-\frac{\omega_{i,\text{M}}}{2}}
					\frac{\partial \omega_{i,\text{M}}}{\partial \gamma_{i,\text{M}}}
					\frac{z_i^2(z_iP_\Sigma+\sigma^2_i)}{(z_iP\subscript{K}+\sigma^2_i)^3}
					\left(
					\omega_{i,\text{M}}
					\frac{\partial \omega_{i,\text{M}}}{\partial \gamma_{i,\text{M}}}
					\underbrace{
						\frac{
							(z_iP_\Sigma+\sigma^2_i)
						}{
							(z_iP\subscript{K}+\sigma^2_i)
						}
					}_{\geqslant \gamma_{i,\text{M}}}
					-2
					\right)\\
					\geqslant&\frac{1}{\sqrt{2\pi}}e^{-\frac{\omega_{i,\text{M}}}{2}}
					\frac{\partial \omega_{i,\text{M}}}{\partial \gamma_{i,\text{M}}}
					\frac{z_i^2(z_iP_\Sigma+\sigma^2_i)}{(z_iP\subscript{K}+\sigma^2_i)^3}
					\left(
					\omega_{i,\text{M}}
					\frac{\partial \omega_{i,\text{M}}}{\partial \gamma_{i,\text{M}}}
					\gamma_{i,\text{M}}-2
					\right)\geqslant\frac{6.25\gamma_{i,\text{M}}}{\sqrt{\gamma_{i,\text{M}}(\gamma_{i,\text{M}}+1)}}-2\geqslant 0,
				\end{split}
			\end{equation}
		\end{strip}
		i.e. $\epsilon_{i,\text{M}}$ is convex of $P\subscript{M}$. The inequality holds with $\gamma_{i,M}\geq \gamma_{th}\geq 1$, which is required to fulfill the error probability constraints in practical scenarios~\cite{Zhu_convex_2022}.
	\end{proof}
	
	\section{Proof of Theorem~\ref{th:concavity_key_length}}
	\begin{proof}
		The second derivative of $U\subscript{FP}$ w.r.t. $d\subscript{K}$ is given by: 
		\begin{equation}
			\begin{split}
				\frac{
					\partial^2 U
				}{
					\partial d^2\subscript{K}
				}
				= &2(1-\epsilon_{\eve,\text{M}})\frac{
					\partial^2 \epsilon_{\eve,\text{K}}
				}{
					\partial d^2\subscript{K}
				}\\
				&+2(1-\epsilon_{\bob,\text{M}})\frac{
					\partial^2 (-\epsilon_{\bob,\text{K}})
				}{
					\partial d^2\subscript{K}
				},
			\end{split}
		\end{equation}
		where
		\begin{equation}
			\frac{
				\partial^2 \epsilon_{\eve,\text{K}}
			}{
				\partial d^2\subscript{K}
			}
			=\frac{1}{mV_{\eve,\text{K}}}
			\underbrace{
				\frac{
					\partial^2 \epsilon_{\eve,\text{K}}
				}{
					\partial \omega^2_{\eve,\text{K}}
				}
			}_{\leqslant 0}
			\leqslant 0,
		\end{equation}
		\begin{equation}
			\frac{
				\partial^2 (-\epsilon_{\bob,\text{K}})
			}{
				\partial d^2\subscript{K}
			}
			=-\frac{1}{mV_{\eve,\text{K}}}
			\underbrace{
				\frac{
					\partial^2 \epsilon_{\bob,\text{K}}
				}{
					\partial \omega^2_{\bob,\text{K}}
				}
			}_{\geqslant 0}
			\leqslant 0.
		\end{equation}
		The above inequalities hold, since we have proven in~\cite{Zhu_convex_2022} that         $\frac{
			\partial^2 \epsilon
		}{
			\partial \omega^2
		}\geqslant 0$, if $\epsilon\leqslant 0.5$ and $\frac{
			\partial^2 \epsilon
		}{
			\partial \omega^2
		}\leqslant 0$, if $\epsilon\geqslant 0.5$. As a result, we have  $    \frac{
			\partial^2 U\subscript{FP}
		}{
			\partial d^2\subscript{K}
		}\leqslant 0$ in the feasible region of Problem \eqref{prob:max_utility_degraded}, which confirms the concavity.
	\end{proof}
	
	\section{Proof of Theorem~\ref{th:concavity_message_power}}\label{app:proof_concavity_message_power}
	\begin{proof}
		We reformulate the utility $U\subscript{FP}$ and group its components as follows:
		\begin{equation}
			\begin{split}
				&U\subscript{FP}=U_\bob-U_\eve\\
				=&(1-\epsilon_{\bob,\text{M}})-2(1-\epsilon_{\bob,\text{M}})\epsilon_{\bob,\text{K}}\\
				 &+2(1-\epsilon_{\eve,\text{M}})\epsilon_{\eve,\text{K}}-(1-\epsilon_{\eve,\text{M}})\\
				=&\underbrace{(\epsilon_{\eve,\text{M}}-\epsilon_{\bob,\text{M}})}_{A_1}
				+
				\underbrace{2(\epsilon_{\eve,\text{M}}-\epsilon_{\bob,\text{M}})(\epsilon_{\eve,\text{M}}+\epsilon_{\bob,\text{M}}-1)}_{A_2}\\
				&+
				2\underbrace{\left[
					(\epsilon_{\eve,\text{K}}-\epsilon_{\bob,\text{K}}) 
					+
					(\epsilon_{\bob,\text{M}}\epsilon_{\bob,\text{K}}-\epsilon_{\eve,\text{M}}\epsilon_{\eve,\text{K}})\right]}_{A_3}
			\end{split}
		\end{equation}
		Note that $U\subscript{FP}$ is concave if each component $A_k$ where $k\in\{1,2,3\}$ are concave.  Therefore, to this end we show the concavity of each $A_k$.
		
		For $A_1$, we have already revealed with Lemma~\ref{lemma:convexity_power} the concavity of both $\epsilon_{i,\text{M}}$ w.r.t. $P\subscript{M}$ for both $i\in\{\bob,\eve\}$.
		Moreover, $\epsilon_{i,\text{M}}$ is also monotonically decreasing in $z_i$ since
		\begin{equation}
			\begin{split}
				\frac{\partial \epsilon_{i,\text{M}}}{\partial z_i}
				&=\frac{\partial \epsilon_{i,\text{M}}}{\partial \gamma_i}
				\frac{\partial \gamma_i}{\partial z_i}=\frac{\partial \epsilon_{i,\text{M}}}{\partial \gamma_i}\frac{P\subscript{M}\sigma^2_i}{(z_ip_m+\sigma^2_i)^2}
				\leqslant 0.
			\end{split}
		\end{equation}
		Note that $\epsilon_{\eve,\text{M}}$ and $\epsilon_{\bob,\text{M}}$ actually distinguish from each other only regarding different values of $z_i$. Since we considerthat  $z_\bob\geqslant z_\eve$, we have $A_1=\epsilon_{\eve,\text{M}}-\epsilon_{\bob,\text{M}}=\epsilon(P\subscript{M},z_\eve)-\epsilon(P\subscript{M},z_\bob)\geqslant 0$ and $\frac{\partial A^2_1}{\partial P\subscript{M}^2}=\frac{\partial^2 \epsilon(P\subscript{M},z_\eve)}{\partial P\subscript{M}^2}-\frac{\partial^2 \epsilon(P\subscript{M},z_\bob)}{\partial P\subscript{M}^2}\geqslant 0$. 
		
		For $A_2$, we have 
		\begin{equation}
			\begin{split}
				\frac{\partial^2 A_2}{\partial P\subscript{M}^2}=&\underbrace{\frac{\partial^2 A_1}{\partial P\subscript{M}^2}}_{\geqslant 0}\underbrace{(\epsilon_{\eve,\text{M}}+\epsilon_{\bob,\text{M}}-1)}_{\leqslant 0}\\
				&+\underbrace{A_1}_{\geqslant 0}\left(\underbrace{\frac{\partial^2 \epsilon_{\eve,\text{M}}}{\partial P\subscript{M}^2}}_{\leqslant 0}+\underbrace{\frac{\partial^2 \epsilon_{\bob,\text{M}}}{\partial P\subscript{M}^2}}_{\leqslant 0}\right)\\
				&+2\underbrace{\frac{\partial A_2}{\partial P\subscript{M}}}_{\geqslant 0}\left(\underbrace{\frac{\partial \epsilon_{\eve,\text{M}}}{\partial P\subscript{M}}}_{\leqslant 0}+\underbrace{\frac{\partial \epsilon_{\bob,\text{M}}}{\partial P\subscript{M}}}_{\leqslant 0}\right)
				\leqslant 0.
			\end{split}
		\end{equation}
		
		For $A_3$, first we note that due to the simple form of $\epsilon_{i,\text{K}}$ given by\eqref{eq:per_key_approx}, all features of $\varepsilon$ regarding $\omega$ and $\gamma$ that we have proven in App.~\ref{app:proof_convexity_power} also hold for $\epsilon_{i,\text{K}}$ with both $i\in\{\bob,\eve\}$. However,	albeit the term $\epsilon_{\eve,\text{K}}-\epsilon_{\bob,\text{K}}$ has the similar structure of $A_1$, the conclusion of concavity cannot be directly applied. This is due to the fact that we have $\epsilon_{\eve,\text{K}}\geqslant \epsilon\superscript{th}_{\eve,\text{K}}\geqslant 0.5$, which implies that $\epsilon_{\eve,\text{K}}$ is actually concave of $\omega_{\eve,\text{K}}$. Moreover, since the mask can only be decoded after SIC, the convexity of $\omega(\gamma_{i,\text{M}})$ w.r.t. $P\subscript{M}$ must also be revisited. In view of this, we first show that the $\epsilon_{\bob,\text{K}}$ is still convex of $P\subscript{M}$ with:
		\begin{equation}
			\begin{split}
				\frac{\partial^2 \epsilon_{\bob,\text{K}}}{\partial P\subscript{M}^2}
				=&  
				\frac{
					\partial^2 \epsilon_{\bob,\text{K}}
				}{
					\partial \gamma^2_{\bob,\text{K}}
				}
				\left(
				\frac{\partial \gamma_{\bob,\text{K}}}{\partial P\subscript{M}}
				\right)^2=
				\frac{z^2_\bob}{\sigma^4_\bob}
				\underbrace{\frac{
						\partial^2 \epsilon_{\bob,\text{K}}
					}{
						\partial \gamma^2_{\bob,\text{K}}
				}}_{\geqslant 0}
				\geqslant 0.
			\end{split}
		\end{equation}
		Whilst, $\epsilon_{\eve,\text{K}}$ is concave of $P\subscript{M}$ since 
		\begin{equation}
			\begin{split}
				\frac{\partial^2 \epsilon_{\eve,\text{K}}}{\partial P\subscript{M}^2}
				=&\frac{\partial^2 \epsilon_{\eve,\text{K}}}{\partial \omega^2_{\eve,\text{K}}}
				\left(\frac{\partial \omega_{\eve,\text{K}}}{\partial \gamma_{\eve,\text{K}}}\right)^2
				+\frac{\partial \epsilon_{\eve,\text{K}}}{\partial \omega_{\eve,\text{K}}}
				\frac{\partial^2 \omega_{\eve,\text{K}}}{\partial \gamma^2_{\eve,\text{K}}}
				\leqslant 0.
			\end{split}
		\end{equation}
		The inverse of the convexity/concavity for $\epsilon_{\eve,\text{K}}$ is due to the fact that we have the constraints $\epsilon_{\eve,\text{K}}\geqslant \epsilon\superscript{th}_{\eve,\text{K}}> 0.5$ and $\epsilon_{\bob,\text{K}}\leqslant \epsilon\superscript{th}_{\bob,\text{K}}< 0.5$, which implies that $\mathcal{C}_{\eve,\text{K}}-d\subscript{K}/n \leqslant 0$ and $\mathcal{C}_{\bob,\text{K}}-d\subscript{K}/n \geqslant 0$.  Finally, combing the above results, for $A_3$ we have:
		\begin{multline}
			\label{eq:A_3}
			\frac{\partial^2 A_4}{\partial P\subscript{M}^2}=
			\underbrace{(\epsilon_{\bob,\text{M}}-1)\frac{\partial^2 \epsilon_{\bob,\text{K}}}{\partial P\subscript{M}^2}}_{\leqslant 0}
			+
			\underbrace{(1-\epsilon_{\eve,\text{M}})\frac{\partial^2 \epsilon_{\eve,\text{K}}}{\partial P\subscript{M}^2}}_{\leqslant 0}\\
			+
			\underbrace{\frac{\partial^2 \epsilon_{\bob,\text{M}}}{\partial P\subscript{M}^2}\epsilon_{\bob,\text{K}}
				-
				\frac{\partial^2 \epsilon_{\eve,\text{M}}}{\partial P\subscript{M}^2}\epsilon_{\eve,\text{K}}}_{B_1}
			\\+
			2
			\underbrace{\left(\frac{\partial \epsilon_{\bob,\text{M}}}{\partial P\subscript{M}}
				\frac{\partial \epsilon_{\bob,\text{K}}}{\partial P\subscript{M}}
				-
				\frac{\partial \epsilon_{\eve,\text{M}}}{\partial P\subscript{M}}
				\frac{\partial \epsilon_{\eve,\text{K}}}{\partial P\subscript{M}}\right)}_{B_2},
		\end{multline}
		where $B_1$ and $B_2$ are auxiliary functions. Recall that $z_\bob\geqslant z_\eve$. Therefore, it holds that $\epsilon_{\bob,\text{K}}\leqslant \epsilon_{\bob,\text{K}}$ regardless of $P\subscript{M}$, with which we can reformulate $B_1$ as
		\begin{equation}
			\label{eq:B_1}
			B_1\leqslant  \left(\frac{\partial^2 \epsilon_{\bob,\text{M}}}{\partial P\subscript{M}^2}-\frac{\partial^2 \epsilon_{\eve,\text{M}}}{\partial P\subscript{M}^2}\right)\epsilon_{\eve,\text{K}}\leqslant 0.
		\end{equation}
		
		Moreover, it also implies that $\gamma_{\eve,\text{M}}\leqslant \gamma_{\bob,\text{M}}$, as well as $\gamma_{\eve,\text{K}}\leqslant \gamma_{\bob,\text{K}}$. In other words, there exist such factors $\alpha\subscript{M}\geqslant 1$ and  $\alpha\subscript{K}\geqslant 1$ that: 
		\begin{align}
			\gamma_{\bob,\text{M}}&=\frac{z_\bob P_c}{z_\bob P_m+\sigma^2}=\frac{\alpha\subscript{M}z_\eve P_c}{z_\eve P_m+\sigma^2}=\alpha\subscript{M}\gamma_{\eve,\text{M}},\\
			\gamma_{\bob,\text{K}}&=\frac{z_\bob P_m}{\sigma^2}=\frac{\alpha\subscript{K}z_\eve P_c}{\sigma^2}=\alpha\subscript{K}\gamma_{\eve,\text{K}}.
		\end{align}
		Therewith, $B_2$ is given by:
		\begin{equation}
			\label{eq:B_2}
			\begin{split}
				B_2=&\frac{\partial \epsilon_{\bob,\text{M}}}{\partial \omega_{\bob,\text{M}}}
				\frac{\partial \omega_{\bob,\text{M}}}{\partial P\subscript{M}}
				\frac{\partial \epsilon_{\bob,\text{K}}}{\partial \omega_{\bob,\text{K}}}
				\frac{\partial \omega_{\bob,\text{K}}}{\partial P\subscript{M}}\\
				&-
				\frac{\partial \epsilon_{\eve,\text{M}}}{\partial \omega_{\eve,\text{M}}}
				\frac{\partial \omega_{\eve,\text{M}}}{\partial P\subscript{M}}
				\frac{\partial \epsilon_{\eve,\text{K}}}{\partial \omega_{\eve,\text{K}}}
				\frac{\partial \omega_{\eve,\text{K}}}{\partial P\subscript{M}}\\
				\leq&
				\underbrace{(\alpha\subscript{K} \alpha\subscript{M}-1)}_{\geqslant 0}
				\underbrace{\frac{\partial \epsilon_{\eve,\text{M}}}{\partial \omega_{\eve,\text{M}}}}_{\leqslant 0}
				\underbrace{\frac{\partial \omega_{\eve,\text{M}}}{\partial P_c}}_{\geqslant 0}
				\underbrace{\frac{\partial \epsilon_{\eve,\text{K}}}{\partial \omega_{\eve,\text{K}}}}_{\leqslant 0}
				\underbrace{\frac{\partial \omega_{\eve,\text{K}}}{\partial P_c}}_{\leqslant 0}\\
				\leqslant&0
			\end{split}
		\end{equation}
		Applying \eqref{eq:B_1} and \eqref{eq:B_2} to \eqref{eq:A_3}, it results in $\frac{\partial^2 A_4}{\partial P\subscript{M}^2}\geqslant 0$.
		
		Now we have proven that each $A_k$, $\forall k=\{1,2,3\}$, is concave of $P\subscript{M}$. Since the sum of concave functions is still concave, we can conclude that $U\subscript{FP}$ is concave in $P\subscript{M}$.
	\end{proof}
\fi



\bibliographystyle{IEEEtran}
\bibliography{references}

\end{document}